\documentclass[prd,showpacs,preprint,nofootinbib]{revtex4}
\usepackage{graphicx,color,amsmath}

\begin{document}
\title{Light scalar mesons in QCD sum rules with inclusion of  instantons}
\author{J. Zhang$^1$, H.Y. Jin$^1$, Z. F. Zhang$^2$, T.G. Steele$^3$ and D. H. Lu$^1$}
\affiliation{$^1$Institute of Modern Physics, Zhejiang University,
Hangzhou, Zhejiang, China\\
$^2$Department of Physics, Ningbo University, Ningbo, Zhejiang, China\\
$^3$Department of Physics and Engineering Physics, University of
Saskatchewan, Saskatoon, Saskatchewan, Canada S7N 5E2}
\date{March 2009}
\begin{abstract}
The light scalar meson nonet above 1~GeV
({\it i.e.\ }the $a_{0}$, $K^{\ast}_{0}$
and $f_{0}$) are studied within the  framework of QCD sum rules. In
conventional QCD sum rules, the calculated masses of this nonet
are degenerate, and  the mass of $K^{\ast}_{0}$
is always larger than the $a_{0}$ in contradiction with the observed
spectrum. After improving the correlation function by
including instanton effects, the masses are well separated from
each other. In particular, our results shows glueball content plays an
important role in the underlying structure of $f_{0}(1500)$. The
decay constants are also discussed.

\end{abstract}


\maketitle
\newpage
\section{introduction}
The $SU(3)$ classification of strongly-interacting particles originally proposed by
Gell-Mann \cite{gell-mann} and Zweig \cite{zweig}\footnote{Following Ref.\, \cite{Amsler 2} we refer to this model as the naive quark model.} has been a very successful paradigm in particle physics.
It is observed from
the hadronic spectrum that the results following from this naive quark
model (NQM) agree better with the heavy meson systems than the light ones. This is
understandable because the heavy quarks inside a heavy meson are non-relativistic, and hence we can deal with their
kinematics in the framework of nonrelativistic quantum mechanics as a
good approximation. However, for  light mesons where  the
light components are relativistic,  it is hard to say whether the nonrelativistic
approximation is applicable. We can see that there is more
complexity in light mesons than the heavy ones from the observed
spectrum. The situation is even worse in combination with the
proliferation of light scalars and their production in charmless $B$
decays. In order to accommodate these light mesons in theories
consistent with QCD, models beyond the naive quark model have been developed, including glueballs \cite{glueball}, multiquark
states \cite{Jaffe 1} and hybrid states \cite{hybrid}. One hopes that these
models can supply some reasonable, or at least qualitative,
explanation of the observed light mesons.

The underlying structure of mesons with mass near
 1 GeV attract much attention. It has widely been suggested
that the light scalars below or near 1GeV [the isoscalars
$f_{0}(600)$, $f_{0}(980)$, the the isodoublet
$K^{\ast}_{0}(800)$(or $\kappa$) and the isovector $a_{0}$(980)] form
a SU(3) flavor nonet, while scalar mesons above 1GeV
[$f_{0}(1370)$, $a_{0}$(1450), $K^{\ast}_{0}$(1430) and
$f_{0}$(1500)/$f_{0}$(1710)] form another nonet \cite{Close 1, Amsler
2, Spanier 3, Godfrey 4}. Refs.~ \cite{Jaffe 1, Delepine 2}
suggest that the light scalar nonet above 1GeV can be
accommodated in the conventional $\bar{q}q$ model with some gluonic
component, while the light scalars around 1GeV are dominated by
$\bar{q}q\bar{q}q$ states with some $0^{+}$ $\bar{q}q$ and glueball states.
But this interpretation is still far from deciphering the puzzle presented by the
light scalars.

It is obvious that the starting point of all the models mentioned
above concentrates on the kinematic aspect of the component inside
the scalars, {\it i.e.,} in order to reproduce the spectrum in
theories consistent with QCD, the complexity of the light scalars is
attributed to their constituents. Maybe one can refer to this as a
kinematics-dependent approach. There is another viewpoint we can
adopt. We should recognize that in the hadronic region  perturbative
QCD breaks down, and the nonperturbative aspects of QCD are
dominant. It is well-known the nonperturbative aspect of QCD is
difficult to analyze. The nonzero quark condensate signals that the
QCD vacuum is nontrivial and has a complex structure. In other word,
the dynamics in QCD vacuum is very different from the trivial one.
It is possible that the complexity of the light scalar mesons can be
attributed to the enigmatic QCD vacuum, in which the particle
treated as the excitation of the QCD vacuum from the viewpoint of
quantum field theory.  The nonzero value of QCD vacuum expectation
values is one of the main ingredients of QCD sum rules \cite{shifman
1, reinders 2, novikov 3} which deals with the low energy
nonperturbative aspects of QCD. In conventional QCD rules the
physical quantities are expressed by a dominant perturbative part
and corrections associated with vacuum expectation values of various
operators. This method works well in many cases, but when we apply
this method to the pions, it is difficult to obtain reasonable
results. This difficulty was solved by introducing instanton
contribution into the QCD sum rules  \cite{shuryak 1}.
Instantons---the nontrivial solution to the Yang-Mills field
equation
  \cite{Belavin}---play an important role in solving the puzzle.\footnote{ For details on instantons in QCD, see
 the excellent review by T. Sch\"{a}fer and E. V. Shuryak
 \cite{sch}}.  Recent work involving QCD sum rules with instanton
effects include the electromagnetic pion form factor \cite{fork 1}
and glueballs \cite{Kisslinger 2}.  Furthermore, instanton effects
within QCD the non-strange sum-rules for scalar currents have
previously  been shown to split the degeneracy between the $a_0$ and
$f_0$ \cite{fang}.\footnote{We note that these works did not
consider the structure of the entire nonet.} All these works pave a
new way to resolving the controversy concerning the nature of the
light scalars.

Comparing with the kinematics-dependent approach, we refer to the
instanton effects as a dynamics-dependent approach, because here one
attempts to solve the problem by further investigating  low energy
QCD itself. Keeping these motivations in mind, in this paper we
investigate the masses of the scalar nonet above
1GeV [i.e, $f_{0}(1370)$, $a_{0}(1450)$, $K^{\ast}_{0}(1430)$ and
$f_{0}(1500)$/$f_{0}(1710)$] from QCD sum rules based on scalar
interpolating fields including the corresponding instanton
contribution. Because there is no mixing between $a_{0}$,
$K_{0}^{\ast}$ meson and the glueball, these two members
are ideally suited
 to investigate the role of instantons in
QCD sum rules, and we will analyze them in the naive quark model.
The situation is more complicated for the $f_0$ because it is widely accepted
 that there is mixing with the
 isoscalar $0^{++}$ glueball ground state  \cite{Amsler 2} around
1500\rm{MeV}

\footnote{Lattice gauge calculations predict a glueball mass of 1400
to 1800 \rm{MeV} \cite{lattice_quench}.} Because
of this mixing, a more consistent analysis should consider the
mixing of quark and gluonic content in analyzing $f_{0}$ meson. So
we will employ a mixed quark-glueball current to
discuss $f_{0}$ meson if necessary. Specifically, we assign
$f_{0}(1500)$ and $f_{0}(1710)$ to be a mixed current of quark and
gluonic content, while $f_{0}(1370)$ is still assumed to be purely
of quark content. As will be demonstrated below, the
validity of these assignments is upheld by the results of the QCD
sum-rule analysis. The instanton contributions to the sum-rules are
calculated using the semiclassical approximation with  quark zero
modes. As a byproduct, the decay constants of the states are
obtained naturally.

In Section II we derive the QCD sum rules with scalar interpolating
fields in the absence of instantons and note the shortcomings
associated with the results of this analysis. In Section III we
present the sum rule including instanton contribution based on
scalar current or its mixing with gluonic current, and the masses
and decay constants of the nonet are calculated. Section IV is
devoted  to our conclusions.

\section{Sum rules without instantons}

In this section we will discuss QCD sum-rules without instantons and
the results following it. The starting point is the following
correlator defined in terms of scalar interpolating current:
\begin{equation}
\Pi(q^2)=i\int d^4 x\,e^{iq \cdot x}\langle0|j(x) j^{\dagger}(0)
|0\rangle,
\end{equation}
where $j(x)$ is a scalar composite operator defined as:
\begin{equation}
j(x)={\bar{q}_{{1}}}(x)q_{2}(x).
\label{current}
\end{equation}
compared with the definition in \cite{shuryak 1}, we have suppressed
the renormalization invariant factor $(\ln(\mu/\Lambda))^{-4/b}$,
with $\mu$ is the normalization pint and $b=(11N_{c}-2n_{f})/3$. The
correlator can be expressed in terms of operator product expansion,
up to order-$\alpha_{s}$ perturbative correction and dimension-six,
the operator product expansion we get is \cite{reinders 2, jamin}:
\begin{eqnarray}
\Pi^{{\rm{OPE}}}(q^{2})&=&-\frac{3}{8\pi^{2}}(1+\frac{11}{3}\frac{\alpha_{s}}{\pi})\
q^{2}\ln
\frac{-q^{2}}{\mu^{2}}+\frac{3}{4\pi^{2}}m_{1}m_{2}\ln\frac{-q^{2}}{\mu^{2}}-\frac{1}{8\pi}
\frac{1}{q^{2}}\langle\alpha_{s}G^{2}\rangle
\nonumber\\
&&-\frac{1}{q^{2}}(\frac{m_{1}}{2}+m_{2})\langle\bar{q}_{1}q_{1}\rangle
-\frac{1}{q^{2}}(\frac{m_{2}}{2}+m_{1})\langle\bar{q}_{2}q_{2}\rangle
\nonumber\\
&&-\frac{1}{2q^{4}}m_{2}\langle g_{s}\bar{q}_{1}\sigma
Gq_{1}\rangle-\frac{1}{2q^{4}}m_{1}\langle g_{s}\bar{q}_{2}\sigma
Gq_{2}\rangle
\nonumber\\
&&-\frac{16\pi}{27}\frac{\alpha_{s}}{q^{4}}\Big[\langle\bar{q}_{1}q_{1}\rangle^{2}+\langle\bar{q}_{2}q_{2}\rangle^{2}\Big]
\nonumber\\
&&-\frac{48\alpha_{s}}{9}\frac{1}{q^{4}}\langle\bar{q}_{1}q_{1}\rangle\langle\bar{q}_{2}q_{2}\rangle
\end{eqnarray}
This is the theoretical side of the QCD sum rule from the quark-gluon
dynamics point of view. On the other hand, the correlator can also
be derived phenomenologically:
\begin{equation}
\Pi(q^2)=\frac{1}{\pi}\int_{0}^{\infty}ds\,
\frac{\rm{Im}\Pi^{ph}(s)}{s-q^{2}} + {\rm subtraction~constants} ~.
\end{equation}
The quantity
$\rm{Im}\Pi^{ph}(s)$ obtained by inserting a complete set of
quantum states $\Sigma |n\rangle \langle n|$ into Eq.~(1), which
reads:
\begin{equation}
{\rm{Im}}\Pi^{\rm{ph}}(q^{2})=m_{S}^{2}f_{S}^{2}\pi\delta(q^{2}-m^{2})+\big[\frac{3}{8\pi^{2}}\pi
(1+\frac{11}{3}\frac{\alpha_{s}}{\pi})q^{2}-\frac{3}{4\pi^{2}}m_{1}m_{2}\pi\big]\theta(q^{2}-s_{0}),
\end{equation}
where $s_0$ represents the onset of the QCD continuum.
The decay constant in Eq.~(5) is defined as:
\begin{equation}
\langle S|\bar{q}_{2}q_{1}|0\rangle=m_{S}f_{S}.\nonumber
\end{equation}
By equating both the theoretical and phenomenological sides, we
obtain the total dispersion integral:
\begin{equation}
\Pi^{{\rm{OPE}}}(q^{2})=\frac{1}{\pi}\int_{0}^{\infty}ds\,
\frac{\rm{Im}\Pi^{ph}(s)}{s-q^{2}} + {\rm substraction~constants } ,
\end{equation}
After Borel transform and subtracting the perturbative continuum
contributions, we obtain the following sum rule:
\begin{eqnarray}
m_{S}^{2}f_{S}^{2}\exp[-\frac{m_{S}^{2}}{M^{2}}]
&=&\int_{0}^{s_{0}}ds\,
\big[\frac{3}{8\pi^{2}}(1+\frac{11}{3}\frac{\alpha_{s}}{\pi})s-\frac{3}{4\pi^{2}}m_{1}m_{2}\big]\exp[-\frac{s}{M^{2}}]+\frac{1}{8\pi}\langle\alpha_{s}G^{2}\rangle
\nonumber\\
&&+(\frac{m_{1}}{2}
+m_{2})\langle\bar{q}_{1}q_{1}\rangle+(m_{1}+\frac{m_{2}}{2})\langle\bar{q}_{2}q_{2}\rangle
\nonumber\\
&&-\frac{1}{2M^{2}}m_{2}\langle g_{s}\bar{q}_{1}\sigma
Gq_{1}\rangle-\frac{1}{2M^{2}}m_{1}\langle g_{s}\bar{q}_{2}\sigma
Gq_{2}\rangle \nonumber\\
&&+\frac{16\pi}{27}\frac{\alpha_{s}}{M^{2}}\Big[\langle\bar{q}_{1}q_{1}\rangle^{2}+\langle\bar{q}_{2}q_{2}\rangle^{2}\Big]
-\frac{48}{9}\frac{\alpha_{s}}{M^{2}}\langle\bar{q}_{1}q_{1}\rangle\langle\bar{q}_{2}q_{2}\rangle.
\end{eqnarray}
where scale dependence of decay constant $f_{S}$ is:
\begin{equation}
f_{S}(M)=f_{S}(\mu)\Big(\frac{\alpha_{s}(\mu)}{\alpha_{s}(M)}\Big)^{4/b},\nonumber
\end{equation}
The parameters in Eq.~(7) are as follows \cite{Khodjamirian 1, Cheng
2}:
\begin{eqnarray}
\alpha_{s}=0.517,
&\langle\frac{\alpha_{s}}{\pi}G^{2}\rangle=0.012\pm0.006\rm{GeV^{4}},
\nonumber\\
\langle\bar{u}u\rangle=\langle\bar{d}d\rangle=-(0.24\pm0.1)^{3}\rm{GeV^{3}},
&\langle\bar{s}s\rangle=(0.8\pm0.2)\langle\bar{u}u\rangle,
\nonumber\\
\frac{m_{u}+m_{d}}{2}=5\rm{MeV}, & m_{s}=120\rm{MeV},
\nonumber\\
\langle g_{s}\bar{u}\sigma Gu\rangle=\langle g_{s}\bar{d}\sigma
Gd\rangle=0.8{\rm{GeV^{2}}}\langle\bar{u}u\rangle,& \langle g_{s}
\bar{s}\sigma Gs\rangle=0.8\langle g_{s}\bar{u}\sigma
Gu\rangle.\nonumber
\end{eqnarray}
All the values adopted here are given at the scale $\mu=1\rm{GeV}$. By
taking the logarithm of both sides of Eq.~(7) and applying the
differential operator $M^4\partial/\partial M^2$ to them, we derive
the desired mass formula which is free of the decay constant.

The task now is to find ranges of parameters $M^{2}$ and the continuum
threshold $s_{0}$ such that the resulting mass does not depend too
much on the value of these parameters. In addition, the continuum
contribution that is the part of dispersive integral from $s_{0}$ to
$\infty$  subtracted from both sides of Eq.~(7) should
not be too large (less than 30\% of the total dispersive
integral), and the contribution of the dimension-six operators is less
than 10\%. One more requirement is the value of the continuum
threshold $s_{0}$ should not stray too much away from the next known
resonance in that channel  \cite{Ball}.

Before proceeding with our analysis we note that
experimentally, except for the $f_{0}(1710)$, there is a small mass
difference between other members, and the mass difference between $a_{0}$
and $K_{0}^\ast$ is even smaller. Thus it is reasonable to deal with them
using the same threshold and Borel window;  we think this criterion
also holds true for other multiplets with small mass
differences considered by other QCD practitioners. Of course one can analyze
each member of a multiplet with a separate threshold and Borel window,
but it is too artificial to be adopted because the sum rules are
sensitive to the threshold. In considering this, we will analyze
this nonet within same threshold and Borel window below.

\begin{figure}
\begin{center}
\includegraphics[scale=1.0]{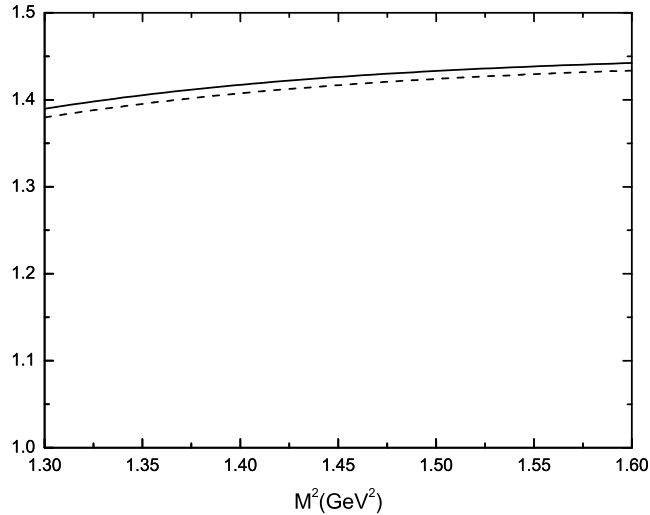}
\caption{Mass of $K_{0}^{\ast}$(solid line) and $a_{0}$ meson
(dashed line) from sum rule Eq.~(7) based on naive quark model as
function of Borel parameter $\rm{M^{2}}$ without instanton.}
\end{center}
\end{figure}

If one uses the same threshold and Borel window, it is easy to see
the mass spectrum of the nonet following from Eq.~(7) will be similar to the naive quark model;
 there is a mass degeneracy broken by a  tiny difference
resulting from the SU(3) flavor symmetry breaking.\footnote{In addition, the
degeneracy could also be broken when mass corrections proportional to
$\alpha_s^2m_{\mu(s)}^2$  are taken into account. These corrections
are negligible compared with instanton effects which we will
consider in the next section.}  Even worse, the mass of
$K_{0}^{\ast}$ with underlying structure $s\bar{d}$ \,is always
larger than the $a_{0}$ with underlying structure
$\frac{1}{\sqrt{2}}(u\bar{u}-d\bar{d})$. For definiteness, when we select
$s_{0}=4.1\rm{GeV^{2}}$ and $M^{2}$ within the range
[1.3,\,1.6]$\rm{GeV^{2}}$,the calculated mass of $K_{0}^{\ast}$ and
$a_{0}$ is shown in Figure\,1. One can see the mass of
the $K_{0}^{\ast}$ is above the $a_{0}$, which is inverted compared to the
experimental results. All the results following from Eq.~(7) are
unsatisfactory.

To summarize Section \uppercase\expandafter{\romannumeral2}, we
conclude that in the conventional QCD sum rule analysis based on the naive quark
model, one can not
separate the this nonet with the same threshold and Borel window,  the masses following from Eq.~(7) are
degenerate,  and the results for the $K_{0}^\ast$ and $a_{0}$,
are in contradiction with  experiment. This suggests that important effects have been neglected in Eq.~(7).

\section{Sum rule with inclusion of instanton contributions}

\subsection{Basic formula}
It has been known for a long time that the instanton plays an
important role in nonperturbative QCD. The starting point on this
subject is the solution of classical field equations in four
dimension Euclidean gauge-field theories given by Belavin \emph{et
al.} \cite{Belavin}. Subsequently t'Hooft derived the instanton with
topological quantum number $n=1$ in Euclidean space \cite{Hooft}:
\begin{eqnarray}
A^{a}_{\mu
}(x)=\frac{2}{g}\eta_{a\mu\nu}\frac{(x-x_{0})_{\nu}}{(x-x_{0})^{2}+\rho^{2}}\,,
\nonumber\\
G^{a}_{\mu\nu}(x)=-\frac{4}{g}\eta_{a\mu\nu}\frac{\rho^{2}}{\big[(x-x_{0})^{2}+\rho^{2}\big]^{2}}\,,
\end{eqnarray}
where $\rho$ is instanton size, $\eta_{a\mu\nu}$ is the t'Hooft
$\eta$ symbol, $x_{0}$ is an any point in Euclidean space. The
density $n(\rho)$ of instanton with size $\rho$ in the vacuum can be
parameterized as  \cite{shuryak 1, Diakonov 2}:
\begin{equation}
n(\rho)=n_{c}\delta(\rho-\rho_{c}),\,
\end{equation}
with two parameters $n_{c}$ and $\rho_{c}$, called the average
instanton density and size:
\begin{equation}
n_{c}=8\times
10^{-4}\rm{GeV^{4}},\quad\rho_{c}=\frac{1}{0.6}\rm{GeV^{-1}}.
\end{equation}
When we include the instanton contribution in the correlator~(1), there is a
new term  \cite{shuryak 1}:
\begin{equation}
\Pi^{\bar{q}q,\,\rm{inst}}(q^{2})=\Bigg|\int d^{4}x\,e^{iq\cdot
x}\bar{q}_{10}(x)q_{20}(x)\Bigg|^{2}\frac{n_{c}}{m_{1}^{\ast}m_{2}^{\ast}},\,
\end{equation}
where $q_{10}$ and $q_{20}$ is the t'Hooft quark zero mode,
respectively, $m_{1}^{\ast}$ and $m_{2}^{\ast}$ is the effective
mass correspondingly. Similarly applying dispersive relation to
Eq.~(11) we can rewrite Eq.~(11) as:
\begin{equation}
\Pi^{\bar{q}q,\,\rm{inst}}(q^{2})=\frac{1}{\pi}\int_{0}^{\infty}ds\,
\frac{{\rm{Im}}\Pi^{\bar{q}q,\,{\rm{inst}}}(s)}{s-q^{2}},\,
\end{equation}
After the Borel transformation we get the desired form of the
instanton contributions of the current with isospin $I$:
\begin{eqnarray}
\Pi^{\bar{q}q,\,{\rm{inst}}}(M^{2})=(-1)^{I}\,\frac{n_{c}\rho^{4}M^{6}}{2m_{1}^{\ast}m_{2}^{\ast}}\exp[-\frac{M^{2}\rho^{2}}{2}]\bigg[K_{0}(\frac{M^{2}\rho^{2}}{2})
+K_{0}(\frac{M^{2}\rho^{2}}{2})\bigg],\,
\end{eqnarray}
and the instanton continuum contribution is:
\begin{equation}
\Pi^{\bar{q}q,\,\rm{inst,\,cont}}(s_{0}, M^{2})=(-1)^{I}\,\frac{\pi
n_{c}\rho^{2}}{m_{1}^{\ast}m_{2}^{\ast}}\int_{s_{0}}^{\infty}ds\,
sJ_{1}(\rho \sqrt{s})Y_{1}(\rho \sqrt{s})e^{-s/M^{2}},
\end{equation}
where $K_{0}$, $K_{0}$ are the McDonald functions, and $J_{1}$ ,
$Y_{1}$ are the Bessel functions.

When the smoke clears, we get the final result:
\begin{eqnarray}
m_{S}^{2}f_{S}^{2}\exp[-m_{S}^{2}/M^{2}]&=&\Pi^{\rm{OPE}}(M^{2})-\Pi^{\rm{OPE,\,
cont}}(s_{0}, M^{2})
\nonumber\\
&&+\Pi^{\bar{q}q,\,\rm{inst}}(M^{2})-\Pi^{\bar{q}q,\,\rm{inst,\,cont}}(s_{0},
M^{2}).
\end{eqnarray}
This is the sum rule we obtained including instanton effects in the
correlation function. Similarly we can obtain the mass from Eq.~(15)
with the same manipulation as the previous section.

\begin{figure}
\begin{center}
\includegraphics[scale=1.0]{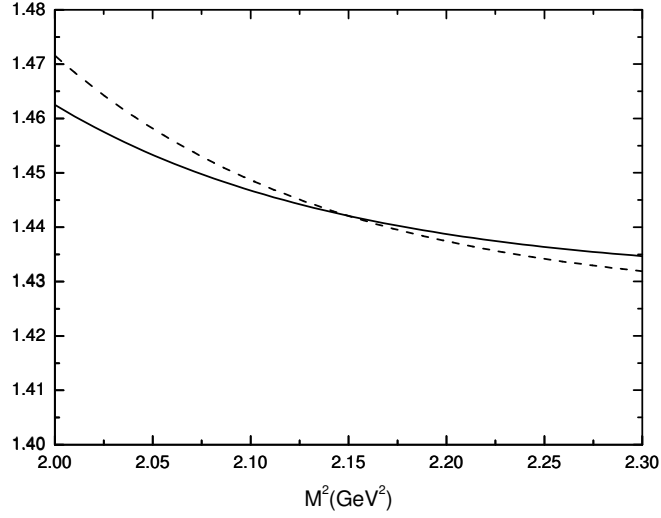}
\caption{Mass of $K_{0}^{\ast}$(solid line) and $a_{0}$ (dashed
line) from sum rule Eq.~(15) as function of Borel parameter
$\rm{M^{2}}$ include instanton }\label{fig1}
\end{center}
\end{figure}

As an important parameter in sum rule Eq.~(15) it is necessary to
discuss the value of the effective mass $m^{\ast}$'s. In the
mean-field approximation  \cite{sch}:
\begin{equation}
m_{u}^{\ast}=m_{d}^{\ast}=\pi\rho\,\bigg(\frac{2}{3}\bigg)^{\frac{1}{2}}(N/V)^{\frac{1}{2}}=170\rm{MeV}\,,
\end{equation}
The value of $N/V$ is $N/V=1{\rm{fm^{-4}}}$ phenomenologically. And
there is a relation between $\langle\bar{u}u\rangle$ and
$\langle\bar{s}s\rangle$:
\begin{equation}
\langle\bar{s}s\rangle=(0.8\pm 0.2)\langle\bar{u}u\rangle\,,
\end{equation}
together with
\begin{equation}
\langle\bar{q}q\rangle=-\frac{N/V}{m_{q}^{\ast}}\,,
\end{equation}
we obtain the effective mass of the strange:
\begin{equation}
m_{s}^{\ast}=215_{-45}^{+68}\rm{MeV}.
\end{equation}
In following we take the central value of \,$m_{s}^{\ast}$,
i.e,\,$m_{s}^{\ast}=220\rm{MeV}$.
\subsection{Mass, decay constant of $K_{0}^{\ast}$ and $a_{0}$ meson}

All the parameters needed in numerical calculation have now been
fixed. Firstly we take $j=\bar{d}s$ which corresponds to the
$K_{0}^{\ast}$ meson as our ``\,sample\,'', since in this case there
is no mixing with glueball, allowing us to examine the effect of instants without the complications
presented by mixing.
.

Using standard QCD sum-rule methodologies, we obtain the threshold
$s_{0}=3.5\rm{GeV^{2}}$, which is just below the next excited state
$K^*(1950)$, and the Borel window is within the range [2.0,
2.3]$\rm{GeV^{2}}$. The calculated mass of $K_{0}^{\ast}$ is
$m_{K_{0}^{\ast}}=1436\sim1462\rm{MeV}$. For the $a_0$, assigning
$j=\frac{1}{\sqrt{2}}(\bar{u}u-\bar{d}d)$ with the same threshold
and Borel window,  the mass of the $a_{0}$ is
$m_{a_{0}}=1432\sim1472\rm{MeV}$. The results for these two states
are shown together in Figure~2.

We can see from Figure\,2 that there is a crossover at the point
$M^{2}=2.16\rm{GeV^{2}}$  corresponding to the mass
$m=1442\rm{MeV}$, which is very close to the experimental value of
$m_{K_{0}^{\ast}}$ and $a_{0}$. We refer to this value of $M^2$ as
the ``key point'' representing the Borel scale where the mass
hierarchy between the $a_0$ and $K_{0}^{\ast}$ reverses. The more
important aspect observed from figure\,2 is that the calculated mass
for $K_{0}^{\ast}$ and $a_{0}$ present the right picture\,: there is
a range where the mass of the $a_{0}$ is larger than the
$K_{0}^{\ast}$, a result which cannot be obtained in the sum rule
without instanton contributions shown in figure\,1. In other words,
the sum rule including instanton contributions can reproduce
realistic results which in agreement with the light meson spectrum.

Having determined the mass, it is straightforward to obtain the
decay constant from Eq.~(15). The decay constants of the $K_{0}^{\ast}$ and $a_{0}$ are shown in
figure\,3 and figure\,4 respectively. It is obvious that the results
are very stable within the Borel window, and from the Figures we find
the decay constants of $K_{0}^{\ast}$ and $a_{0}$:
\begin{eqnarray}
f_{K_{0}^{\ast}(1430)}(\rm{1GeV})=510MeV, &f_{a_{0}(1450)}(\rm{1GeV}
)=514MeV\nonumber
\end{eqnarray}

\begin{figure}
\begin{center}
\includegraphics[scale=1.0]{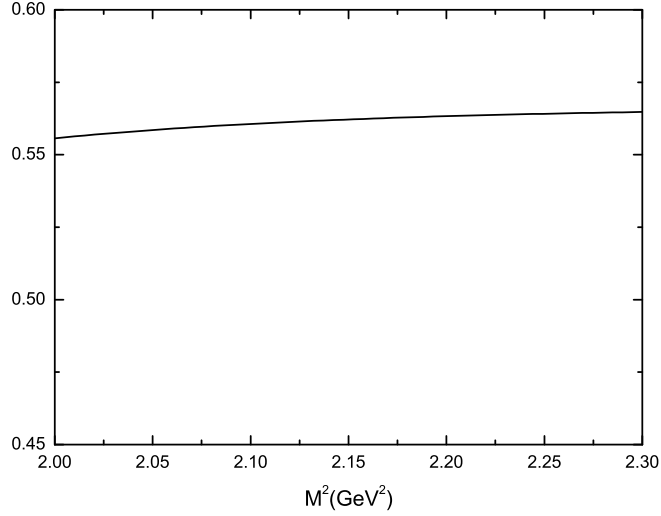}
\caption{Decay constants of $K_{0}^{\ast}$ with quark structure
$s\bar{d}$ as function of Borel parameter $\rm{M^{2}}$ includes
instanton.}
\end{center}
\end{figure}

\begin{figure}
\begin{center}
\includegraphics[scale=1.0]{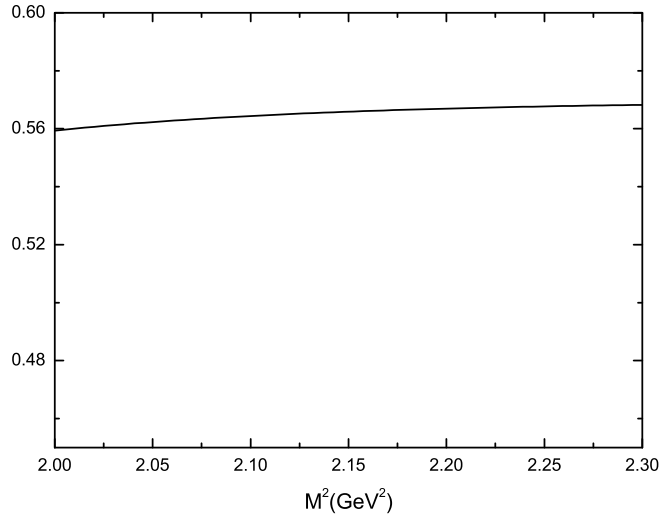}
\caption{Decay constants of $a_{0}$ with quark structure
$(u\bar{u}-d\bar{d})/\sqrt{2}$ as function of Borel parameter
$\rm{M^{2}}$ includes instanton.}
\end{center}
\end{figure}

\subsection{Mass, decay constant of $f_{0}$ meson with underlying structure $\frac{1}{\sqrt{2}}(u\bar{u}+d\bar{d})$}
Similarly if we set
\begin{equation}
j=\frac{1}{\sqrt{2}}(\bar{u}u+\bar{d}d),\nonumber
\end{equation}
with isospin $I=0$, we immediately get the mass of $f_{0}$\,:
\begin{equation}
m_{f_{0}}=1314\sim1391\rm{MeV},\nonumber
\end{equation}
At the ``\,key point\,'' i.e, $M^2=2.16\rm{GeV^{2}}$, the mass is
$m_{f_{0}}=1376\rm{MeV}$ which is close to the experimental value of
$f_{0}(1370)$, while the more important result is that the mass of
this state is no longer degenerate with other states. The decay
constant following this mass is:
\begin{equation}
f_{f_{0}(1370)}(\rm{1GeV})=520\rm{MeV}.\nonumber
\end{equation}
The mass and decay constant of $f_{0}$ are shown in figure\,5 and
figure\,6, respectively.

Because there are more controversies for the $f_{0}$ meson than the two
members discussed above, it is useful to mention the model beyond
pure quark viewpoint. We also notice the glueball can mix with
scalar mesons nearby, so it is possible that there is mixing between
$f_{0}(1370)$ glueball, or in other words, $f_{0}(1370)$ is not in a
pure quark state, and could have some glue content; an
 idea that was first introduced in Ref. \cite{camsler} and then
generalised in Ref. \cite{feclose}. The work in Ref. \cite{feclose}
suggested the $f_{0}(1710)$ is dominated by $s\bar{s}$ content,
while $f_{0}(1500)$ and $f_{0}(1370)$ share roughly equal amounts
glueball($\simeq 40\%$) ($f_{0}(1500)$ and $f_{0}(1710)$ will be
detailed in the coming subsection). Here we see when including the
instanton effects, $f_{0}(1370)$ can be accommodated naturally in
pure quark model. So we conclude that $f_{0}(1370)$ may be a pure quark
state based when instanton effects are considered.

\begin{figure}
\begin{center}
\includegraphics[scale=1.0]{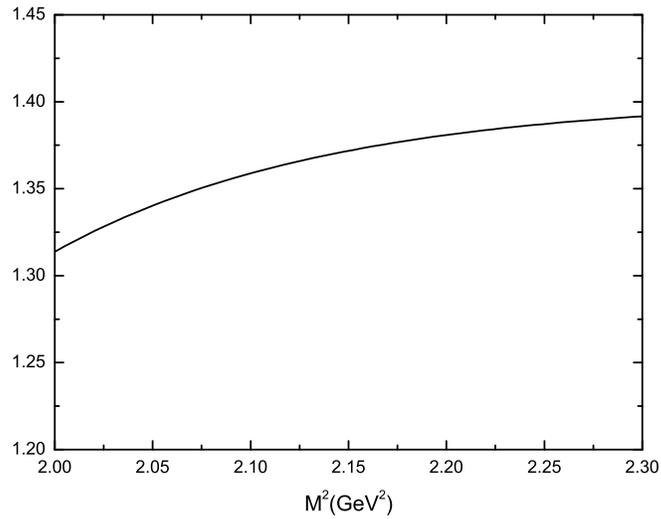}
\caption{Mass of $f_{0}$ with quark structure\,
$(u\bar{u}+d\bar{d})/\sqrt{2}$\, as function of Borel parameter
$\rm{M^{2}}$ includes instanton.}
\end{center}
\end{figure}

\begin{figure}
\begin{center}
\includegraphics[scale=1.0]{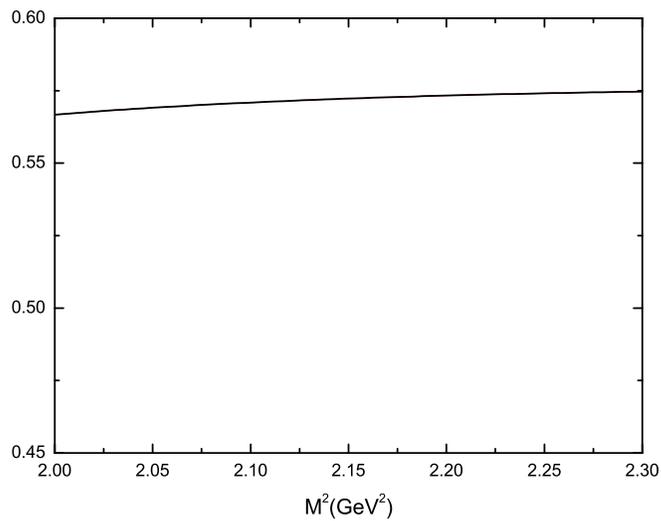}
\caption{Decay constants of $f_{0}$  with quark structure\,
$(u\bar{u}+d\bar{d})/\sqrt{2}$\, as function of Borel parameter
$\rm{M^{2}}$ includes instanton.}
\end{center}
\end{figure}

\subsection{Mass, decay constant of $f_{0}$ meson with underlying
structure $s\bar{s}$}

The $f_{0}(1500)$ may be the most controversial object in this
nonet. As we have seen the important role of the instanton in giving
the mass of the three members in previous subsections. Firstly we do
some tentative calculations based on pure quark model in hope that
these calculations shed some light on its structure. In order to
present a thorough investigation on this object, it is reasonable to
write the current with isospin $I=0$ in a general form:
\begin{equation}
j=c_{1}(\bar{u}u+\bar{d}d)+c_{2}\bar{s}s,
\end{equation}
which includes  two adjustable parameters $c_{1}$ and $c_{2}$. In
case of $c_{1}=0$, Eq.~(20) reduces to pure $s\bar{s}$ state which
is free of instanton \cite{shuryak 1}  so we can deal it with the conventional
QCD sum rule as in section II. The calculated mass is shown in
figure\,7 from which we can read the mass with pure $s\bar{s}$
structure:
\begin{equation}
m_{f_{0}}=1413\sim1430\rm{MeV},\nonumber
\end{equation}

\begin{figure}
\begin{center}
\includegraphics[scale=1.0]{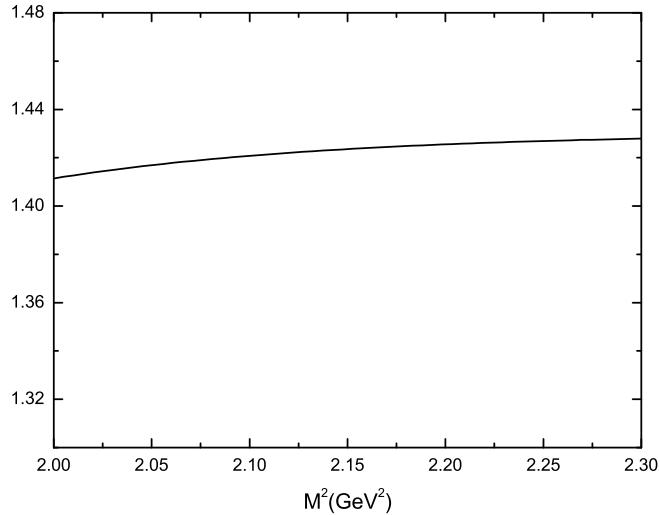}
\caption{Mass of $f_{0}$ with quark structure\, $s\bar{s}$ as
function of Borel parameter $\rm{M^{2}}$ without instanton.}
\end{center}
\end{figure}

We can also write the current in a more complex form as:
\begin{equation}
j=\frac{1}{\sqrt{6}}\Big[(\bar{u}u+\bar{d}d)-2\bar{s}s\Big],\nonumber
\end{equation}
such that there will be instanton effects in the corresponding sum
rule. The calculated mass with this quark content is shown in
figure\,8 from which the following mass is read:
\begin{equation}
m_{f_{0}}=1433\sim1457\rm{MeV},\nonumber
\end{equation}

In fact, based on the current given in Eq.~(20), we can derive a sum
rule involving complete instanton contributions induced by the
quarks which depends on the two adjustable parameters $c_{1}$ and
$c_{2}$. Unfortunately the results indicate the sum rule is not able
to produce  reasonable masses for the $f_{0}(1500)$ and $f_{0}(1710)$ by
adjusting these two parameters---the masses are always much lower than
the experimental ones.

\begin{figure}
\begin{center}
\includegraphics[scale=1.0]{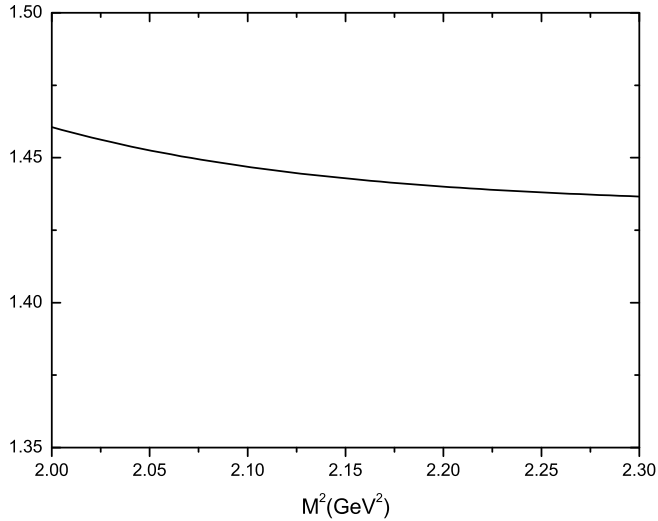}
\caption{Mass of $f_{0}$ with quark structure\,
$\big[(\bar{u}u+\bar{d}d)-2s\bar{s}\big]/\sqrt{6}$\, as function of
Borel parameter $\rm{M^{2}}$ includes instanton contributions.}
\end{center}
\end{figure}

These simple calculations signal that from the viewpoint of pure
quark content, one can not produce the $f_{0}(1500)$ in QCD sum
rules even including complete instanton contributions, so we must
look for other solution to this problem. As mentioned in
Ref. \cite{Amsler 2}, if we assume a $q\bar{q}$ structure, one
concludes that $f_{0}(1500)$ is dominantly $s\bar{s}$, while this
assignment can not produce reasonable mass theoretically as we can
see from previous paragraphs, but also leads to contradictions
experimentally \cite{Amsler 2}. There are some works \cite{Kisslinger
2} on this subject that take  another extreme: they try to produce
$f_{0}(1500)$ under the assumption of a pure glueball content. But
what is the realistic structure of $f_{0}(1500)$ is still unknown.

There is another viewpoint that the light nonet above 1\rm{GeV} can
be identified as conventional $\bar{q}q$ states with some possible
gluonic content, that is, there is mixing of the pure glueball with
the nearby two $N=n\bar{n}$ and $S=s\bar{s}$ scalar mesons as first
introduced in Ref. \cite{camsler}, where
$n\bar{n}=\frac{1}{2}(u\bar{u}+d\bar{d})$. Based on this model, Ref.
\cite{feclose} obtained the results that $f_{0}(1710)$ is dominated
by $s\bar{s}$ content while there is roughly equal amounts of glue
content in $f_{0}(1500)$. We have seen the key role of instanton in
solving the puzzle on $K_{0}^{\ast}$ and $a_{0}$, $f_{0}(1370)$, and
explore this possibility in the assumed mixing of scalar meson and
pure glueball in $f_{0}(1500)$. With this motivation, we modified
the current of $f_{0}(1500)$ as mixing of quark and gluonic
current:\footnote{ The renormalization-group invariant gluonic
current has been used because the subleading perturbative effects
will be included in the correlation function.}
\begin{equation}
j_{mix}=A\bar{s}s+B\alpha_{s}G_{\mu\nu}^{a}G^{a\mu\nu},
\end{equation}
 and in this case the decay constant is defined as:
\begin{equation}
\langle S|j_{mix}|0\rangle=m_{S}^{2}f_{S}.\nonumber
\end{equation}
 where $A$, $B$ are both
real, and one should notice that the parameter $A$ has dimension one
of mass which insures the right dimension in the
current. The parameters $A$ and $B$
accompany the Wilson coefficients of operators
$\bar ss$ and $\alpha_{s}G_{\mu\nu}^{a}G^{a\mu\nu}$ respectively,
and are therefore renormalization scale dependent. Here we  fix the
renormalization scale of $A$ and $B$ so that they just are numbers
in the following consideration.  After this modification, there will
be new contributions stemming from the glueball
$\alpha_{s}G_{\mu\nu}^{a}G^{a\mu\nu}$ OPE, the glueball instanton
and the mixing instanton contribution which will be presented below.

With the perturbative corrections and including nonperturbative
terms up to dimension eight, the OPE of gluonic current
is \cite{bagan 1, dharnett}:
\begin{eqnarray}
\Pi^{\rm{GB,\,OPE}}(q^{2})&=&q^{4}\ln\frac{-q^{2}}{\mu^{2}}\Bigg\{-2\bigg(\frac{\alpha_{s}}{\pi}\bigg)^{2}\bigg[1+\frac{659}{36}\frac{\alpha_{s}}{\pi}
+247.48\bigg(\frac{\alpha_{s}}{\pi}\bigg)^{2}\bigg]
\nonumber\\
&&+2\bigg(\frac{\alpha_{s}}{\pi}\bigg)^{3}\bigg(\frac{9}{4}+65.781\frac{\alpha_{s}}{\pi}\bigg)\ln\frac{-q^{2}}{\mu^{2}}
-10.125\bigg(\frac{\alpha_{s}}{\pi}\bigg)^{4}\ln^{2}\frac{-q^{2}}{\mu^{2}}\Bigg\}
\nonumber\\
&&+\bigg[4\pi\frac{\alpha_{s}}{\pi}\bigg(1+\frac{175}{36}\frac{\alpha_{s}}{\pi}\bigg)
-9\pi\bigg(\frac{\alpha_{s}}{\pi}\bigg)^{2}\ln\frac{-q^{2}}{\mu^{2}}\bigg]\langle\alpha_{s}G^{2}\rangle
\nonumber\\
&&-8\pi^{2}\bigg(\frac{\alpha_{s}}{\pi}\bigg)^{2}\frac{1}{q^{2}}\langle\mathcal
{O}_{6}\rangle+8\pi^{2}\frac{\alpha_{s}}{\pi}\frac{1}{q^{4}}\langle\mathcal
{O}_{8}\rangle,
\end{eqnarray}
where
\begin{equation}
\langle\mathcal {O}_{6}\rangle=\langle
g_{s}f_{abc}G^{a}_{\mu\nu}G^{b}_{\nu\rho}G^{c}_{\rho\mu}\rangle=(0.27{\rm{GeV^{2}}})\langle\alpha_{s}G^{2}\rangle,\nonumber
\end{equation}
and
\begin{equation}
\langle\mathcal {O}_{8}\rangle=
14\langle(\alpha_{s}f_{abc}G^{a}_{\mu\nu}G^{b}_{\nu\rho})^{2}\rangle
-\langle(\alpha_{s}f_{abc}G^{a}_{\mu\nu}G^{b}_{\rho\lambda})^{2}\rangle=\frac{9}{16}(\langle\alpha_{s}G^{2}\rangle)^{2}.\nonumber
\end{equation}
are the dimension-6 and dimension-8 gluonic condensates,
respectively. Because there is both quark and gluon current, we
have to use the unsubtracted dispersive relation for the gluonic
correlation function in order to be consistent with of the whole
correlation function. Applying the dispersion relation, and after
subtracting the continuum contribution and taking the Borel transform,
the glueball contribution is obtained\, \cite{dharnett, harnett}:
\begin{eqnarray}
\Pi^{\rm{GB,\,OPE}}(s_{0},M^{2})&=&\int_{0}^{s_{0}}ds\,s^{2}e^{-\frac{s}{M^{2}}}\Bigg\{2\bigg(\frac{\alpha_{s}}{\pi}\bigg)^{2}
\bigg[1+\frac{659}{36}\frac{\alpha_{s}}{\pi}+247.48\bigg(\frac{\alpha_{s}}{\pi}\bigg)^{2}\bigg]
\nonumber\\
&&-4\bigg(\frac{\alpha_{s}}{\pi}\bigg)^{3}\bigg(\frac{9}{4}+65.781\frac{\alpha_{s}}{\pi}\bigg)\ln\frac{s}{\mu^{2}}
\nonumber\\
&&-10.125\bigg(\frac{\alpha_{s}}{\pi}\bigg)^{4}\bigg(\pi^{2}-3\ln^{2}\frac{s}{\mu^{2}}\bigg)\Bigg\}
\nonumber\\
&&+9\pi\bigg(\frac{\alpha_{s}}{\pi}\bigg)^{2}\langle\alpha_{s}G^{2}\rangle\int_{0}^{s_{0}}ds\,e^{-\frac{s}{M^{2}}}
\nonumber\\
&&+8\pi^{2}\bigg(\frac{\alpha_{s}}{\pi}\bigg)^{2}\langle\mathcal
{O}_{6}\rangle-8\pi^{2}\frac{\alpha_{s}}{\pi}\frac{1}{M^{2}}\langle\mathcal
{O}_{8}\rangle,
\end{eqnarray}
The contribution of glueball instanton after subtracting continuum
is given by\, \cite{hforkel, dharnett}:
\begin{equation}
\Pi^{\rm{GB,\,inst}}(s_{0},M^{2})=-2^{4}\pi^{3}n\rho_{c}\int_{0}^{s_{0}}ds\,e^{-\frac{s}{M^{2}}}s^{2}J_{2}(\rho\sqrt{s})Y_{2}(\rho\sqrt{s}),
\end{equation}
where $J_{2}$ and $Y_{2}$ are Bessel and Neumann functions,
respectively.

\begin{figure}
\begin{center}
\includegraphics[scale=1.0]{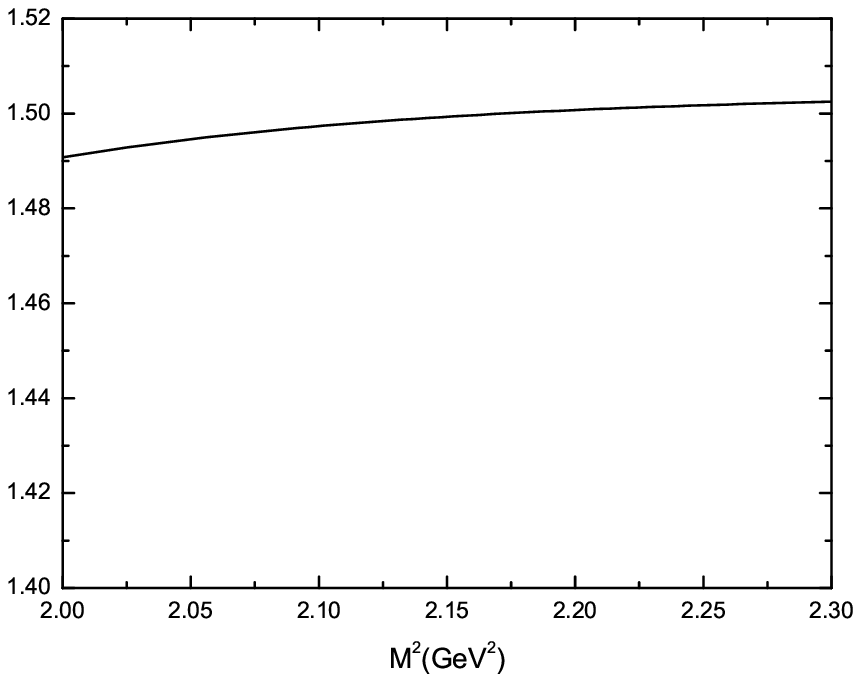}
\caption{Mass of $f_{0}$ with structure
0.9GeV$s\bar{s}+\alpha_{s}G_{\mu\nu}^{a}G^{a\mu\nu}$
 as function of Borel parameter $\rm{M^{2}}$ include glueball instanton and
 mixing instanton contributions.}
\end{center}
\end{figure}

We have independently verified the following instanton contribution
to the mixed correlator
$\bar{s}s\alpha_{s}G^{a}_{\mu\nu}G^{a\mu\nu}$ \cite{harnett}:
\begin{equation}
\Pi^{\rm{mix,\,inst}}(s_{0},M^{2})=\frac{2\pi^{2}n\rho^{3}}{m_{s}^{\ast}}\int_{0}^{s_{0}}ds\,e^{-\frac{s}{M^{2}}}
s^{\frac{3}{2}}\bigg[J_{1}(\rho\sqrt{s})Y_{2}(\rho\sqrt{s})+Y_{1}(\rho\sqrt{s})J_{2}(\rho\sqrt{s})\bigg].
\end{equation}
Now we have determined all the terms induced by the current given by
Eq.~(28). It is convenient to write the whole results in a compact
form as follows:
\begin{equation}
m_{S}^{4}f_{S}^{2}\exp[-\frac{m_{S}^{2}}{M^{2}}]=\sum_{X}\Pi^{X}(s_{0},
M^{2}),
\end{equation}
where $X$ denotes
\begin{equation}
X=\bigg\{\{\bar{s}s,\,\rm{OPE}\}, \rm{\{GB, OPE\}, \{GB,\,inst\},
\{mix,inst\}}\bigg\}.\nonumber
\end{equation}
and we have absorbed the two parameters $A$ and $B$ in the
$\Pi^{X}$'s for convenience. Taking the same algorithm as the
previous section one can obtain immediately the mass corresponding
to the current given in Eq.~(21). Assigning $A=0.9\rm{GeV}$ and
$B=1$ in Eq.~(21), corresponding to a large glueball content (since
the energy scale is $\sim 1\rm{GeV}$), the calculated mass with this
underlying structure is $m_{f_{0}}=1492\sim1504\rm{MeV}$ which is
very close to the experimental result $f_{0}(1500)$.

\begin{figure}
\begin{center}
\includegraphics[scale=1.0]{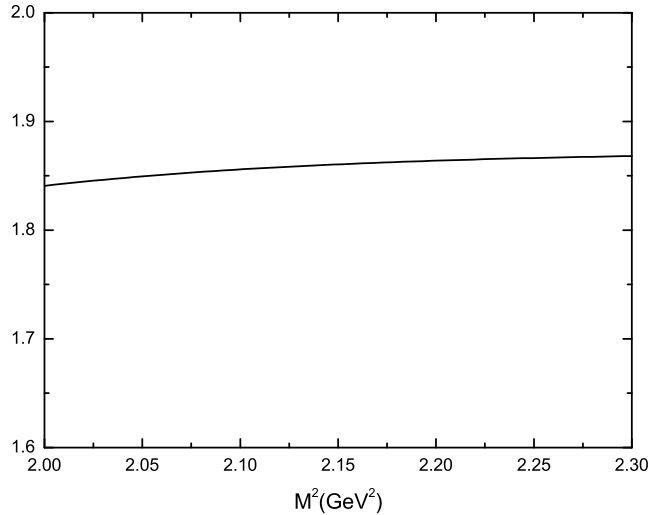}
\caption{Decay constant of $f_{0}$ with structure
0.9GeV$s\bar{s}+\alpha_{s}G_{\mu\nu}^{a}G^{a\mu\nu}$
 as function of Borel parameter $\rm{M^{2}}$ include glueball instanton and
 mixing instanton contributions.}
\end{center}
\end{figure}

After obtaining the mass, we can deduce the decay constant from
Eq.~(26)
\begin{equation}
f_{f_{0}(1500)}(\rm 1GeV)=1.69\rm{GeV}.\nonumber
\end{equation}
There is a strong enhancement of the decay constant after involving
glueball and related instanton contributions compared with other
multiplets. A similarly enhancement of $f_{G}$ in pure glueball
state including instanton effects was found in instanton vacuum
model calculation\, \cite{tshcaefer}. This strong enhancement also
observed in the work of H. Forkel\, \cite{hforkel} which gave a
value $f_{G}=1.14\rm{GeV}$ when the glueball instanton contribution
was included in the pure glueball correlation function using a
unsubtracted dispersive relation. The results for our analysis of
the mass and decay constant of this mixed state are shown in
figure\,9 and figure\,10.

Finally we turn to the last state $f_{0}(1710)$. Generally it is
assumed this state dominated by the $s\bar{s}$ content, so we can
write the current as:
\begin{equation}
j=A'\bar{s}s+B'\alpha_{s}G^{a}_{\mu\nu}G^{a\mu\nu},
\end{equation}
subjected the following orthogonality condition:
\begin{equation}
\langle 0|j|f_0(1500)\rangle=0.\nonumber
\end{equation}
This orthogonal condition is insignificant here since we are not
able to get a value agreeable with $f_{0}(1710)$ whatever the values
of $A'$ and $B'$ are chosen . This can be understood intuitively
that the threshold $s_0=(1.9\rm{GeV})^2$, which is adopted here only
for the states with the masses around 1450GeV,  is too low to
reproduce such a large mass.

\section{Conclusions}

In this work we have studied the mass and decay constant of the light
nonet $a_{0}$, $K_{0}^{\ast}$, and $f_{0}$ within the framework of
QCD sum rule with and without instanton contributions. Our main
results are as follows:
\newline
1.\quad In the conventional QCD sum rule, the masses of this nonet
are degenerate, the calculated mass of $K_{0}^{\ast}$ is larger than the
$a_{0}$ for  the same threshold and same Borel window.
\newline
2.\quad When we include instanton contributions in the sum rule, the
masses of the nonet can be well separated, and the mass of
$K_{0}^{\ast}$ and $a_{0}$ agrees well with the observed results.
The results suggest the underlying structure:
$K_{0}^{\ast}(1430)$  is $s\bar{d}$, $a_{0}(1450)$ is
$\frac{1}{\sqrt{2}}(u\bar{u}-d\bar{d})$, and $f_{0}(1370)$ is
$\frac{1}{\sqrt{2}}(u\bar{u}+d\bar{d})$. For the $f_0(1500)$, our results suggest there
is considerable glueball content in its underlying structure. The
decay constant of $f_{0}(1500)$ enhanced considerably after this
gluonic-content improvement compared with other multiplets.
\newline
3.\quad With a mixing current and the threshold and Borel window common to the multiplet,  we cannot obtain the mass of $f_{0}(1710)$. One reason might be
that the threshold suitable for $K^*(1430)$ is too low for
$f_{0}(1710)$.

\section{Acknowledgements}
This work is supported by NNSFC under Projects No. 10775117,
10847148. The author H.Y. Jin  also thanks KITPC for its
hospitality.  TGS is grateful for research support from the Natural
Sciences and Engineering Research Council of Canada (NSERC).

\end{document}